\documentclass[aps,pra,reprint,groupedaddress,showpacs,9pt,a4paper]{revtex4-1}

\usepackage{graphicx}
\usepackage{braket}
\usepackage{amsbsy}
\usepackage{amsmath}
\begin{document}

\title{Importance of complex orbitals in calculating the self-interaction corrected 
ground state of atoms}


\author{Simon Kl\"{u}pfel}
\email[]{simon@theochem.org}
\affiliation{Science Institute and Faculty of Science, VR-III, University of 
Iceland, Reykjav\'{i}k, Iceland}
\author{Peter Kl\"{u}pfel}
\affiliation{Science Institute and Faculty of Science, VR-III, University of 
Iceland, Reykjav\'{i}k, Iceland}
\author{Hannes J\'{o}nsson}
\affiliation{Science Institute and Faculty of Science, VR-III, University of 
Iceland, Reykjav\'{i}k, Iceland}


\date{\today}
\begin{abstract}
The ground state of atoms from H to Ar was calculated using a self-interaction 
correction to local and gradient dependent density functionals. The correction 
can significantly improve the total energy and makes the orbital energies consistent 
with ionization energies. 
However, when the calculation is restricted to real orbitals, application of the 
self-interaction correction can give significantly higher total energy and worse results, 
as illustrated by the case of the Perdew-Burke-Ernzerhof gradient dependent functional.
This illustrates the importance of 
using complex orbitals for systems described by orbital density dependent energy 
functionals.\end{abstract}


\maketitle


Density functional theory (DFT) of electronic systems has become a widely used tool 
in calculations for solids, liquids, and molecules \cite{Koh99}.
The most commonly used approximation to the exact energy functional for extended systems 
is the so-called generalized gradient approximation (GGA), 
where the functional only depends on the total spin density and its gradient, an 
improvement on the local density approximation (LDA), where gradients are not 
included. The kinetic energy is commonly evaluated by introducing orthonormal 
orbitals, $\varphi_i({\bf r})$, consistent with a given
total electron density,  $\rho({\bf r}) = \sum_i |\varphi_i({\bf r})|^2$.
In many cases, DFT significantly improves the total energy over the Hartree-Fock (HF) 
method and gives acceptable accuracy with smaller computational effort.
However, a number of shortcomings are also known: 
The bond energy tends to be too large while the activation energy for atomic 
rearrangements tends to be underestimated \cite{Nac96}. There is also a tendency to 
over delocalize spin density, sometimes making localized electronic defects 
unstable with respect to delocalization \cite{Pac00}. 
For finite systems, an ionization energy (IE) can be determined as the energy 
difference between the ground states of charged and neutral species. These values 
often agree well with experiment, but ionization from deeper energy levels and 
ionization from solids cannot be estimated this way.
Unlike in HF theory, the energy associated with the orbitals (single particle 
eigenvalues) is in practice neither a good nor a theoretically justified estimate 
of ionization energy. Even the exact DFT functional would give an estimate 
of only the first ionization energy.
 
A leading source of error is the spurious self-interaction introduced when the 
Hartree energy $E_{\text{H}}$ is evaluated from the total electron density 
$\rho({\bf r})$:
\begin{eqnarray}
E_{\rm H}[\rho] = \frac{1}{2}\int\!d^3{\bf r}d^3{\bf r}' \ \frac{\rho(
{\bf r})\rho({\bf r}')}{|{\bf r}-{\bf r}'|}
\label{eq:hartree}
\end{eqnarray}
If the orbital densities $\rho_{i}({\bf r})=|\varphi_i({\bf r})|^2$ represent 
single particle distributions, a more accurate expression is 
\begin{eqnarray}
E^{\text{ODD}}_{\rm H}[\rho^N] = \frac{1}{2}\sum_{i\neq j}\int\!d^3{\bf r}d^3
{\bf r}' \ \frac{\rho_i({\bf r})\rho_j({\bf r}')}{|{\bf r}-{\bf r}'|}.
\label{eq:odd-hartree}
\end{eqnarray}
Here, $\rho^N$ denotes the set of orbital densities $\lbrace \rho_1, \dots, 
\rho_N\rbrace$
corresponding to the set of occupied orbitals, $\lbrace \varphi_1, \dots, 
\varphi_N\rbrace$, denoted by $\varphi^N$.
This expression for the energy is explicitly orbital-density dependent (ODD).
The difference between the two expressions is the diagonal terms ($i=j$), which can 
be interpreted as the interaction of the electron in each orbital with itself.
In Hartree-Fock calculations, where the Hartree energy is evaluated as in Eq.\ 
(\ref{eq:hartree}), the exchange energy includes equally large self-interaction 
with opposite sign, so the self-interaction cancels out exactly. In LDA and GGA 
[collectively referred to as Kohn-Sham (KS) here], the exchange-correlation energy, 
$E_{\text{xc}}$, is approximate and the cancellation is incomplete.
Perdew and Zunger \cite{Per81} proposed an orbital-based self-interaction 
correction (SIC) 
\begin{equation}
E_{\text{SIC}}\left[\rho^N\right]=E_{\text{KS}}\left[\rho\right] - \sum_{i=1}^{N} 
E_{\text{SI}}\left[\rho_i\right]
\label{eq:pzsic}
\end{equation}
using $E_{\text{SI}}\left[\rho_i\right]=E_{\text{H}}\left[\rho_i\right] + 
E_{\text{xc}}\left[\rho_i\right]$.
Other estimates of SIC can be formulated \cite{Leg02andLun01,Vyd06}, 
but here we take Eq.\ (\ref{eq:pzsic}) to be the definition. 
Originally, SIC was proposed for LDA, but it can in principle be applied to other 
functionals.
These are examples of a more general class 
of functionals, ODD functionals, where the Hartree energy is evaluated from 
Eq.\ (\ref{eq:odd-hartree}).

Variational optimization of orbital dependent functionals is typically carried out 
by adding the orthonormality constraints multiplied by Lagrange multipliers, 
$\lambda_{ji}$, to the energy functional to give
\begin{equation}
S
\left[\rho^N\right]=E_{\text{SIC}}\left[\rho^N\right] - \sum_{i,j=1}^{N} 
\lambda_{ji}\left(\braket{\varphi_{i}|\varphi_{j}}-\delta_{ij}\right) 
\end{equation}
and finding a stationary point
with respect to variation of each orbital $|\varphi_i \rangle$ and its complex 
conjugate. 
This gives two sets of equations for the {\it optimal} orbitals \cite{Ped84and85, 
Sva96andGoe97, Mes08and09}, 
\begin{eqnarray}
\label{eq:min_cond_h}
\widehat{H}_i |\varphi_i\rangle  
&=& 
\sum_{i=1}^{N} \lambda_{ji} |\varphi_j\rangle \quad {\rm and} \ \ \   
\boldsymbol{\lambda}  = \boldsymbol{\lambda}^\dagger,
  \label{eq:min_cond_l}
\label{eq:min_cond}
\end{eqnarray} 
with $\widehat{H}_i |\varphi_i\rangle = {\delta E_{\text{SIC}}}/{\delta 
\varphi_i}$ and  
$\lambda_{ji} = \braket{\varphi_j|\widehat{H}_i|\varphi_i}$. The ODD functional 
form leads to orbital-dependent Hamiltonians, $\widehat{H}_i$. 
In the case of SIC, $E_{\text{SI}}$ is the orbital density dependent part of the 
energy while $E_{\text{KS}}$ only depends on the total spin density, so that  
\begin{equation}
 \widehat{H}_i = \widehat{H}_{\text{KS}}[\rho] + {\hat v}[\rho_{i}]
\end{equation} 
where ${\hat v}[\rho_{i}] = -{\delta E_{\text{SI}}\left[\rho_i\right]} / {\delta 
\rho_i} = -({\hat v}_{\text H}[\rho_i] + {\hat v}_{\text xc}[\rho_i])$. 
At the minimum, the constraint matrix $\boldsymbol\lambda$ is Hermitian and can be 
diagonalized
to give orbital energies $\varepsilon_{i}$  and corresponding eigenfunctions, the 
{\it canonical} orbitals, $ \psi^N$. In terms of these, 
condition (\ref{eq:min_cond}) can be written as
\begin{equation}
\widehat{H} |\psi_i\rangle 
= \varepsilon_i |\psi_i\rangle
\ , \quad
\varepsilon_i\delta_{ij} = ({\bf W}{\boldsymbol\lambda}{\bf W^\dagger})_{ij}
\label{eq:canonified}
\end{equation}
with $|\psi_i\rangle= \sum_{k=1}^{N} W_{ki}|\varphi_k\rangle$. The non-local 
operator $\widehat{H}$ is defined in terms of the $N$ optimal orbitals $\varphi^N$ 
and their corresponding Hamiltonians $\widehat{H}_{i}$. In this way, the 
single-particle equations (\ref{eq:min_cond}) can be decoupled to give traditional 
eigenvalue equations \cite{Mes08and09}.  The calculation is carried out using two 
sets of orbitals, $\varphi^N$ and $\psi^N$, while keeping track of the 
transformation matrix, ${\bf W}$, relating the two sets.
In HF and KS-DFT, the energy is independent of the unitary transformation and the 
optimal orbitals are typically chosen to be the same as the canonical orbitals, i.e.,
${\bf W}={\bf 1}$. 

At intermediate steps of the variational optimization of ODD functionals,
$\boldsymbol{\lambda}$ is in general not Hermitian, but can be made Hermitian by 
finding the unitary transformation that zeros the matrix 
$\boldsymbol{\kappa}=\boldsymbol{\lambda}-\boldsymbol{\lambda}^\dagger$ 
\cite{Ped84and85},
where 
\begin{eqnarray}\label{eq:local_cond}
\kappa_{ij}
&=&
\int\!d^3{\bf r} \ 
\varphi^*_i({\bf r})\varphi_j({\bf r})
\left(
v[\rho_{i}]({\bf r}) - v[\rho_{j}]({\bf r})
\right)
= 
0 .
\end{eqnarray}
With a Hermitian $\boldsymbol{\lambda}$ at each iteration, Eq.\ (\ref{eq:canonified}) 
can be used during the minimization of the energy 
in a scheme analogous to what is done in KS-DFT and HF.

The results presented here were obtained with the program \textsc{quantice} \cite{quantice} 
using the Cartesian representation of the Gaussian-type correlation-consistent polarized 
valence quadruple zeta (cc-
pVQZ)\cite{Woo94} basis set and
numerical grids of 75 radial shells of 302 points \cite{Mur96andLeb99}. The 
convergence criterion in the optimization was a squared residual norm below 
10$^{-5}$~eV$^2$.  For LDA, Slater exchange and the Perdew-Wang parameterization 
of correlation is used (SPW92)\cite{LDA}.

\begin{figure}
\centering
\includegraphics{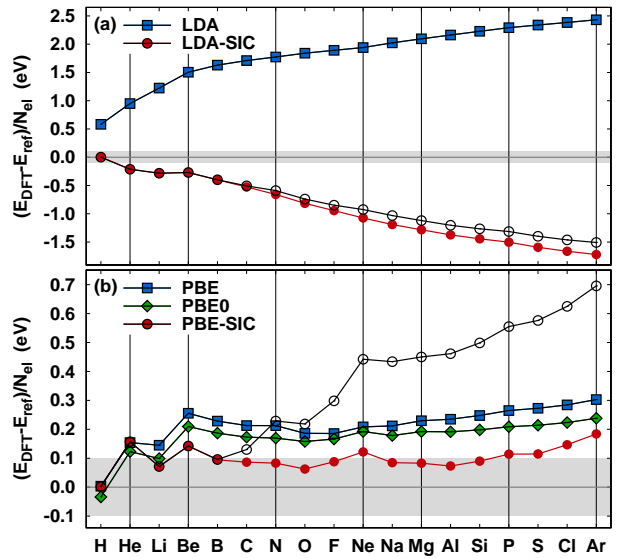}
\caption{(Color online) Energy of atoms H to Ar using LDA, LDA-SIC, PBE, 
PBE-SIC, and PBE0 functionals compared with accurate, non-relativistic estimates 
\cite{Cha96}, normalized to the number of electrons. Results obtained with complex 
orbitals (solid symbols) and real orbitals (open symbols) are also compared. 
Vertical lines indicate the $s^2$, $p^3$, and $p^6$ electron configurations. The 
gray shading at $\pm$0.1~eV illustrates the different energy scales in (a) and (b). 
When complex orbitals are used, PBE-SIC gives significant improvement in the total 
energy, but not when real orbitals are used. 
\label{fig:xc_energy}}
\end{figure}

Figure \ref{fig:xc_energy} compares the energy of the atoms H to Ar, calculated 
using various functionals, with accurate reference values \cite{Cha96}. While the 
inclusion of gradients in the GGA type PBE functional \cite{Per96a}  improves on 
the results obtained with LDA, the energy is still significantly too high and the 
error per electron tends to increase with the size of the atom. SIC applied to LDA 
reduces the magnitude of the error but gives a strong overcorrection. The 
overcorrection also increases with the atomic number. 
When SIC is  applied to the gradient-dependent PBE functional, the error is reduced 
to $\sim$0.1 eV per electron.
But, this improvement is only obtained if the optimal orbitals are complex, i.e., 
complex linear combination coefficients are used for the expansion of the 
orbitals in the Gaussian basis. When real expansion coefficients are used, the 
SIC actually increases the error substantially, as had already been noted 
previously \cite{Vyd04}. 
A common approach to improve the results obtained with DFT is to mix in HF 
exchange with GGA and LDA in so called hybrid functionals \cite{Bec93a}. 
The PBE0 \cite{PBE0} hybrid functional only gives 
slightly better results than PBE (see Fig.\ \ref{fig:xc_energy}). 
PBE-SIC with complex orbitals gives more accurate total energy.

The ionization energy can be evaluated as the energy difference of the charged and 
neutral system. These values typically agree to within $\pm5\%$ with experiment 
for the atoms H to Ar, both for PBE and PBE-SIC. However, in extended systems, 
subject to periodic boundaries, the charged system can not be calculated 
rigorously, so the IE has to be extracted from ground-state properties. 
\begin{figure}
\includegraphics{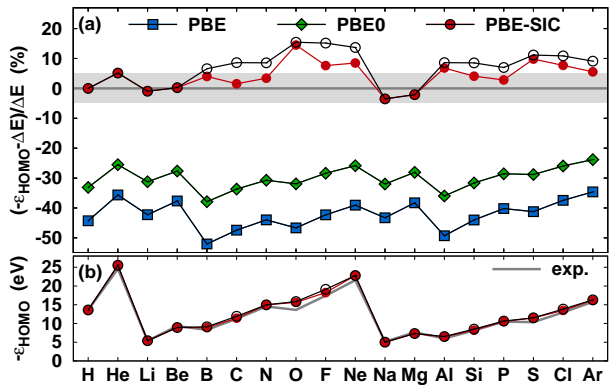}
\caption{\label{fig:ionization_energy}(Color online) 
(a) Comparison of the eigenvalue of the highest occupied orbital 
$\varepsilon_{\text{HOMO}}$ with $\Delta E=E_{\text{cation}}-E_{\text{neutral}}$. 
(b) Comparison of PBE-SIC eigenvalues with experimental ionization energy 
\cite{NIST}. Open and solid symbols are used as in Fig.\ \ref{fig:xc_energy}. 
PBE-SIC eigenvalues agree much better than PBE and PBE0 with both the calculated 
energy difference and experimental data.}
\end{figure}
In Fig.\ \ref{fig:ionization_energy}, the energy eigenvalue of the highest occupied 
canonical 
orbital is compared with the calculated IE using the functionals PBE, PBE0, and
PBE-SIC. As is well known, for the commonly used LDA and GGA functionals, the 
eigenvalues obtained from KS-DFT do not give good estimates of ionization energy.
The IE estimates from PBE and PBE0 eigenvalues are 
too low, with errors of $\sim$40\% and 30\% respectively. In PBE-SIC, however, the 
values are better, in most cases the error is within 5\%-10\%. The 
eigenvalues are also in good agreement with experimental data on IE \cite{NIST}, as shown in Fig.\ 
\ref{fig:ionization_energy}. The values obtained 
from complex orbitals are closer to both experiment and the calculated energy 
difference than those obtained using real orbitals.
A similar improvement is observed for ionization from lower lying orbitals. 
For argon, for example, 
the second, third, and fourth highest orbital energies in PBE-SIC using complex orbitals 
deviate by 5\%, 3\%, and 1\% (1.5, 7, and -3~eV), respectively, from measured 
photoionization energy \cite{Shi77}. Real orbitals give similar deviations of 10\%, 3\%, and 1\% 
(2.8, 7, and 2~eV), while the PBE values are off by 18\%, 8\%, and 10\% (-5.2, -19, 
and -32~eV).

\begin{figure}
\includegraphics{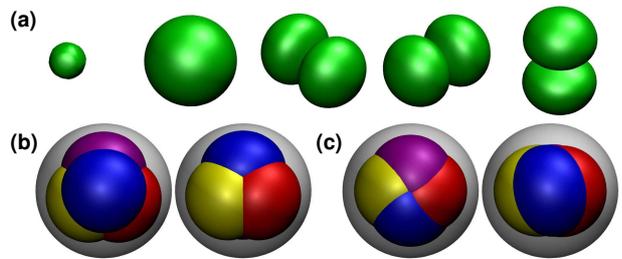}
\caption{(Color online) Orbital densities of a Ne atom obtained from a PBE-SIC 
calculation.
(a) Electron density isosurfaces of the complex canonical orbitals show clear 
correspondence with $s$ and $p$ orbitals. The real canonical orbitals have a similar 
shape. Electron density isosurfaces from (b) real and (c) complex optimal valence 
orbitals shown from the top and side views for one spin component. 
The gray spheres show the total density.}
\label{fig:orb_density2}
\end{figure}

\begin{figure}
\includegraphics{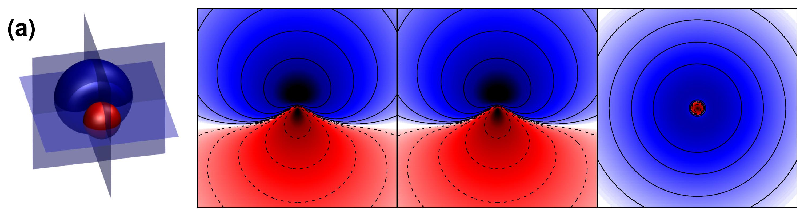}\\%
\includegraphics{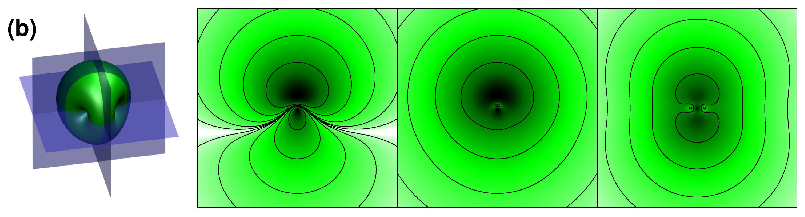}%
\caption{(Color online) Isosurface and contour plots of an optimal valence orbital 
of neon obtained with PBE-SIC 
using (a) real and (b) complex expansion coefficients. The contour plots show the 
orbital density in three planes through the nucleus. The complex orbital has lower 
symmetry and no nodal surface.
\label{fig:orb_density}}
\end{figure}

Figure \ref{fig:orb_density2} compares the PBE-SIC ground state for neon 
using real and complex orbitals. In both cases the canonical orbitals are 
of $s$ and $p$ type and the total density is spherical. The optimal orbitals, however, 
differ significantly in shape. The real orbitals are aligned in the traditional 
tetrahedral $sp^3$ configuration and can be constructed from the 
canonical orbitals as $\varphi_{r1}=\frac{1}{2}(s+p_x+p_y+p_z)$ followed by three 
consecutive fourfold improper rotations about the $z$ axis, 
$\hat{S}_{4}(z)$. The complex optimal orbitals are also, despite their uncommon 
tetragonal configuration, sp$^3$-hybrid orbitals. Such a set can be 
constructed from $\varphi_{c1}=\frac{1}{2}(s+p_x+p_y+{\rm i}\;p_z)$ and 
application of the rotations $\hat{S}_{4}(z)$. The shape of a real and 
complex orbital is compared in Fig.\ \ref{fig:orb_density}. The density of the real 
orbital has axial symmetry and a nodal surface. The complex orbital has lower 
symmetry and lacks the nodal surface since orthogonality is achieved by the phase 
of the complex expansion coefficients.

\begin{figure}
\includegraphics{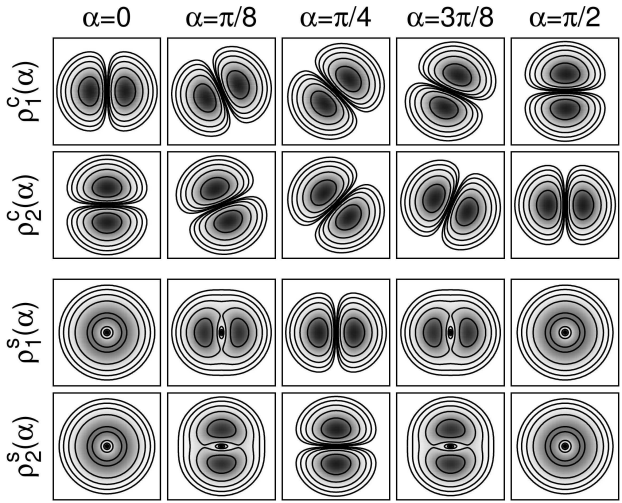}
\caption{Orbital densities obtained from the Cartesian (upper panel) and spherical 
basis (lower panel) using real expansion coefficients. The Cartesian basis only 
gives localized orbitals but allows for all orientations. The spherical basis 
makes a transition from delocalized to localized orbitals possible but is 
restricted to certain orientations of the localized orbitals.
\label{fig:orb_example}}
\end{figure}

The large increase in total energy that occurs when the SIC calculation is 
restricted to real orbitals was also observed for other GGA-SIC functionals
as well as for exchange-only SIC calculations. This effect can be 
explained by the fundamentally different structure of the functional as compared to 
KS or HF.
There, the energy is invariant with respect 
to unitary transformations of the orbitals so the full flexibility of complex 
expansion coefficients is not needed. The SIC energy, however, depends explicitly 
on the orbital densities. 
Nodal surfaces, which are required for orthogonality of real orbitals, impose a 
strong constraint on the shape of the orbital densities (and their gradient, 
resulting in a stronger effect on the energy for SIC applied to GGA functionals 
\cite{Vyd06}).
This can be illustrated by a simple example:
Given a basis set consisting of two Cartesian $p$-type orbitals,
\begin{subequations}
\begin{equation}
\varphi^c_1({\bf r}) =  Nxe^{-\beta{\bf r}^2}
,\ 
\varphi^c_2({\bf r}) =  Nye^{-\beta{\bf r}^2},
\label {Cartesian}
\end{equation}
a second set can be constructed as the complex spherical representation,
\begin{eqnarray}
\varphi^s_1({\bf r}) &=& \frac{N}{\sqrt{2}}\overline{r}e^{{\rm i}\phi} 
e^{-\beta{\bf r}^2}
,\ 
\varphi^s_2({\bf r}) = \frac{N}{\sqrt{2}}\overline{r}e^{-{\rm i}\phi} e^{-\beta
{\bf r}^2},
\label {spherical}
\end{eqnarray}
\end{subequations}
where $\overline{r} = \sqrt{x^2+y^2}$. The two sets of orthogonal orbitals 
accessible by real linear combinations are defined by a single parameter $\alpha$:
\begin{subequations}
\begin{eqnarray}
\widetilde\varphi^x_1 &=& \quad\cos(\alpha)\varphi^x_1 + \sin(\alpha)\varphi^x_2
\\
\widetilde\varphi^x_2 &=& -\sin(\alpha)\varphi^x_1 + \cos(\alpha)\varphi^x_2
\ , \quad x\in\{c,s\}
\end{eqnarray}
\end{subequations}
The orbital densities are illustrated in Fig.\ \ref{fig:orb_example} for several 
values of $\alpha$.
For the Cartesian basis (\ref{Cartesian}), the variation of $\alpha$ results in a 
rotation of the orbitals about the $z$ axis, while for the spherical basis a 
transition from delocalized to localized orbitals occurs. The SIC calculated from 
the Cartesian basis is independent of $\alpha$, but for the spherical basis the 
energy of the localized and delocalized orbitals will differ. While the spherical 
basis (\ref{spherical}) can produce delocalized orbitals, it is incapable of 
describing localized orbital densities pointing in the ``diagonal'' orientation if 
real linear combination coefficients are used.
Only when complex coefficients are used, can both basis sets give the full range of 
possibilities. 

While errors in the total energy of atoms can, to a large extent, cancel out when 
the energy differences of two atomic configurations are calculated, as, for example, 
in calculations of bond energy, we find that the calculated binding energy in diatomic 
molecules is significantly affected both by the inclusion of SIC and then also by a 
restriction to real orbitals.
The binding energy of N$_2$ is predicted by PBE to be $\sim$0.65~eV too large 
compared with experimental estimates (see the references in Ref.\ \cite{Per96a}). PBE-SIC 
reduces the bond energy by 0.69~eV, giving good agreement, while a calculation 
confined to real orbitals overcorrects, giving a bond that is 0.43~eV too weak. For 
O$_2$, PBE predicts a binding energy that is 1.02~eV too large. Here, PBE-SIC 
overcorrects and gives a value that is 0.54~eV too small, while a calculation 
restricted to real orbitals gives an even larger overcorrection, a binding energy that 
is 1.05~eV too small.

The calculations presented here for atoms demonstrate that PBE-SIC gives 
substantial improvement in the total energy and physically meaningful orbital 
energies.
The SIC applied to KS functionals represents only a small set of possible orbital 
density dependent functionals. It is likely that other functionals of ODD form 
allow for an even more accurate modeling of the electronic ground state providing a 
more flexible and computationally efficient alternative to hybrid functionals.
The derivation of a functional that makes optimal use of the more general ODD 
form appears to be a promising prospect.
But, it is clear that an assessment of the quality of any ODD functional 
requires a minimization in the variational space of complex orbitals.


\end{document}